\documentclass[conference]{IEEEtran}
\usepackage[pdftex]{graphicx}
\usepackage{amsmath}
\usepackage{amssymb}
\graphicspath{{images/}}

\begin{document}
\title{Ultrasound Domain Adaptation Using Frequency Domain Analysis}

\author{\
\IEEEauthorblockN{Mostafa~Sharifzadeh, Ali K. Z. Tehrani, Habib~Benali, and Hassan~Rivaz}
\IEEEauthorblockA{Department of Electrical and Computer Engineering\\Concordia University\\ Montreal, QC, Canada\\mostafa.sharifzadeh@concordia.ca}}
\maketitle

\begin{abstract}
A common issue in exploiting simulated ultrasound data for training neural networks is the domain shift problem, where the trained models on synthetic data are not generalizable to clinical data. Recently, Fourier Domain Adaptation (FDA) has been proposed in the field of computer vision to tackle the domain shift problem by replacing the magnitude of the low-frequency spectrum of a synthetic sample (source) with a real sample (target). This method is attractive in ultrasound imaging given that two important differences between synthetic and real ultrasound data are caused by unknown values of attenuation and speed of sound (SOS) in real tissues. Attenuation leads to slow variations in the amplitude of the B-mode image, and SOS mismatch creates aberration and subsequent blurring. As such, both domain shifts cause differences in the low-frequency components of the envelope data, which are replaced in the proposed method. We demonstrate that applying the FDA method to the synthetic data, simulated by Field II, obtains an 3.5\% higher Dice similarity coefficient for a breast lesion segmentation task.
\end{abstract}

\IEEEpeerreviewmaketitle

\section{Introduction}
The performance of deep neural networks (DNNs) is highly correlated with the amount of available data for training, where a larger training set leads to better results. Therefore, insufficient data have always been a concern particularly in the medical domain, where clinical data are not easily accessible due to multiple concerns, and the performance of these methods is hindered by this issue. Using synthetic data for training networks is a popular approach to address this challenge. However, in most cases, the trained models on synthetic data are not well generalizable in real-world applications, where we need to deal with real imaging data acquired by various scanners and different protocols \cite{Ghafoorian2017}.

This issue originally comes from the fact that DNNs typically assume that both training and test sets have been drawn from the same distribution \cite{Chen2018}, which is not necessarily true, especially regarding the recent trend of using synthetic data for training. This problem is usually referred to as the domain shift problem, which induces a dramatic performance drop \cite{Gopalan2011}.

Domain adaptation methods are a well-known solution to address the domain shift problem and have been investigated in the medical ultrasound domain.
Tierney \textit{et al.} proposed a scheme that incorporates both simulated and unlabeled \textit{in vivo} data to train a beamformer. They employed cycle-consistent generative adversarial networks to map between simulated and \textit{in vivo} data in both the input and ground truth target domains \cite{Tierney2020, Tierney2021}.
Ying \textit{et al.} introduced a multi-scale self-attention unsupervised network for domain adaptation between labeled thyroid ultrasound images and unlabeled ones in a different domain \cite{Ying2020}.
Meng \textit{et al.} introduced mutual information-based disentangled neural networks for classifying unseen categories of fetal ultrasound images in different domains. They extracted generalizable categorical features by explicitly disentangling categorical and domain features via mutual information minimization to transfer knowledge to unseen categories in a target domain \cite{Meng2021b}.
Zhang \textit{et al.} proposed a deep-stacked transformation approach for generalizing medical image segmentation models to unseen domains and evaluated it on segmentation tasks involving MRI and ultrasound modalities \cite{Zhang2020a}.

Recently, Yang \textit{et al.} \cite{Yang2020a} introduced the Fourier Domain Adaptation (FDA) method in the field of computer vision. They proposed that moving sample A from the source distribution to the distribution of sample B in the target dataset can be achieved by computing the Fast Fourier Transform (FFT) of both samples and substituting the magnitude of the low-frequency spectrum of the source sample with the target sample and finally reconstructing the modified source sample using inverse FFT (IFFT).
This method is much faster than DNN-based methods, and they demonstrated promising results to adapt synthetic dataset GTA5 \cite{Richter2016} to the real domain dataset CityScapes \cite{Cordts2016}, which both contain urban street scenes.

We believe this method can perform even better on ultrasound images than urban street scenes because two common differences between synthetic and real ultrasound data are caused by unknown values of attenuation and speed of sound (SOS) in real tissues. Attenuation leads to slow variations in the amplitude of the B-mode image, and a mismatch between the nominal and true values of the SOS creates aberration and subsequent blurring. As such, both of these domain shifts are low-frequency in nature and can be compensated by swapping the low-frequency spectrum of the synthetic and real image.

In this work, for the first time, we exploit the FDA method to mitigate the domain shift problem of synthetic ultrasound images and evaluate its performance in a breast lesion segmentation task.

\section{Methodology}
	Let $I\in \mathbb{R}^{W\times H}$ and $\hat{S}\in \{0,1\}^{W\times H}$ denote a sample input image and the corresponding output segmentation mask, respectively, where $W$ and $H$ are width and height of the image. The segmentation problem can be formulated as
\begin{flalign}
\hat{S} = f_{seg}(I, \boldsymbol{\theta})
\end{flalign}
\noindent where $f_{seg}: \mathbb{R}^{W\times H} \rightarrow \{0,1\}^{W\times H}$ is the segmentation convolutional neural network (CNN), and $\boldsymbol{\theta}$ are the network's parameters. By training the CNN, an optimizer is utilized to find optimal parameters $\boldsymbol{\theta^{\ast}}$ that minimize the error, measured by a loss function $L$, between predicted mask $\hat{S}$ and ground truth $S$
\begin{flalign}
\boldsymbol{\theta^{\ast}} = \underset{\theta}{argmin} \; L(S, \hat{S})
\end{flalign}

\subsection{Datasets}
\subsubsection{Synthetic Dataset}
We simulated 1000 ultrasound images using the publicly available Field II simulation package \cite{Jensen1996, Jensen1992} containing 100,000 scatterers uniformly distributed inside a phantom of size 50 mm $\times$ 10 mm $\times$ 50 mm in $x$, $y$, and $z$ directions, respectively. All phantoms were centered at the focal point, positioned at an axial depth of 20 mm from the face of the transducer, and each contained an anechoic region with a different shape. To generate those anechoic regions, we took 1000 samples with only one salient object from a publicly available dataset, denoted as XPIE \cite{Xia2017}, which contained segmented natural images. Then we discarded natural images and resampled only their ground truth masks with the same size as the phantom. Finally, we assigned a zero weight to the amplitude of scatterers which were located inside the mask. The advantages of this method were twofold: First, the mask could be considered as the ground truth of the simulated images. Second, we provided the network with an extended range of features as opposed to regions with limited shapes. Finally, we resampled all images to a size of 256$\times$256 and split the dataset into two training and validation sets, each containing 800 and 200 images, respectively. Note that this data was only used for training and validation and did not contain a test set.
The simulation parameters are summarized in Table~\ref{tbl1}.
\begin{table}[h!]
	\caption{Field II parameters for data simulation.}
	\label{tbl1}
	\setlength{\tabcolsep}{3pt}
	\def\arraystretch{1.5}%
	\begin{tabular}{p{115pt}p{100pt}}
		\hline
		\textbf{Parameter}& 
		\textbf{Value}\\
		\hline
		Center Frequency&
		3.5 MHz\\
		Total/Active Number of Elements& 
		192/64\\
		Element Height& 
		5 mm\\
		Element Width& 
		Equals to wavelength\\
		Kerf& 
		0.05 mm\\
		\hline
	\end{tabular}
\end{table}
\subsubsection{\textit{In vivo} Dataset}
We exploited an ultrasound breast images dataset, known as Dataset B \cite{Yap2018a}. The dataset was publicly available and collected in 2012 from the UDIAT Diagnostic Centre with a Siemens ACUSON Sequoia C512 system and a 17L5 HD linear array transducer. It included 163 breast B-mode ultrasound images containing lesions of different sizes at different locations, with a mean image size of 760$\times$570 pixels. Lesions were categorized into benign and cancerous classes, with 110 and 53 samples in each class, respectively. The dataset also contained respective ground truth masks of the breast lesions, manually obtained by experienced radiologists. We resampled all images to a size of 256$\times$256, and split the dataset into three training, validation, and test sets, each containing 20, 20, and 123 images, respectively.

\subsection{Fourier Domain Adaptation}
To mitigate the domain shift problem, the FDA method \cite{Yang2020a} suggests replacing the magnitude of the low-frequency spectrum of source samples with target samples.
Let $I_s\in \mathbb{R}^{W\times H}$, and $I_t\in \mathbb{R}^{W\times H}$ represent a simulated image using Field II (source) and an \textit{in vivo} image (target), respectively. Besides, let $\mathcal{F}_M(I): \mathbb{R}^{W\times H} \rightarrow \mathbb{R}^{W\times H}$ and $\mathcal{F}_P(I): \mathbb{R}^{W\times H} \rightarrow \mathbb{R}^{W\times H}$ be the magnitude and phase of the Fourier transform $\mathcal{F}$ of the image $I$:
\begin{flalign}
\mathcal{F}(I)(m,n)=\sum_{w=0}^{W-1}\sum_{h=0}^{H-1}I(w,h)e^{\textstyle -j2\pi(\frac{h}{H}n+\frac{w}{W}m)}
\label{eq1}
\end{flalign}

\noindent Accordingly, given $\mathcal{F}_M(I)$ and $\mathcal{F}_P(I)$, $\mathcal{F}^{-1}$ is the inverse Fourier transform that converts back the signal from the frequency domain to the image domain.
\begin{flalign}
I = \mathcal{F}^{-1}(\mathcal{F}_M(I), \mathcal{F}_P(I))
\label{eq2}
\end{flalign}

\noindent where $j^2=-1$, and (\ref{eq1}) and (\ref{eq2}) can be implemented using FFT \cite{Frigo1998} and IFFT algorithms, respectively.

\noindent Further, let denote with $M_\alpha$ a mask matrix of size $W \times H$:
\begin{flalign}
M_\alpha(w,h) =
\begin{cases} 
1, & -\alpha<\frac{2w}{W}-1<\alpha, -\alpha<\frac{2h}{H}-1<\alpha\\
0, & otherwise
\end{cases}
\end{flalign}

\begin{figure*}
	\centering
	\includegraphics[width=0.9999\linewidth]{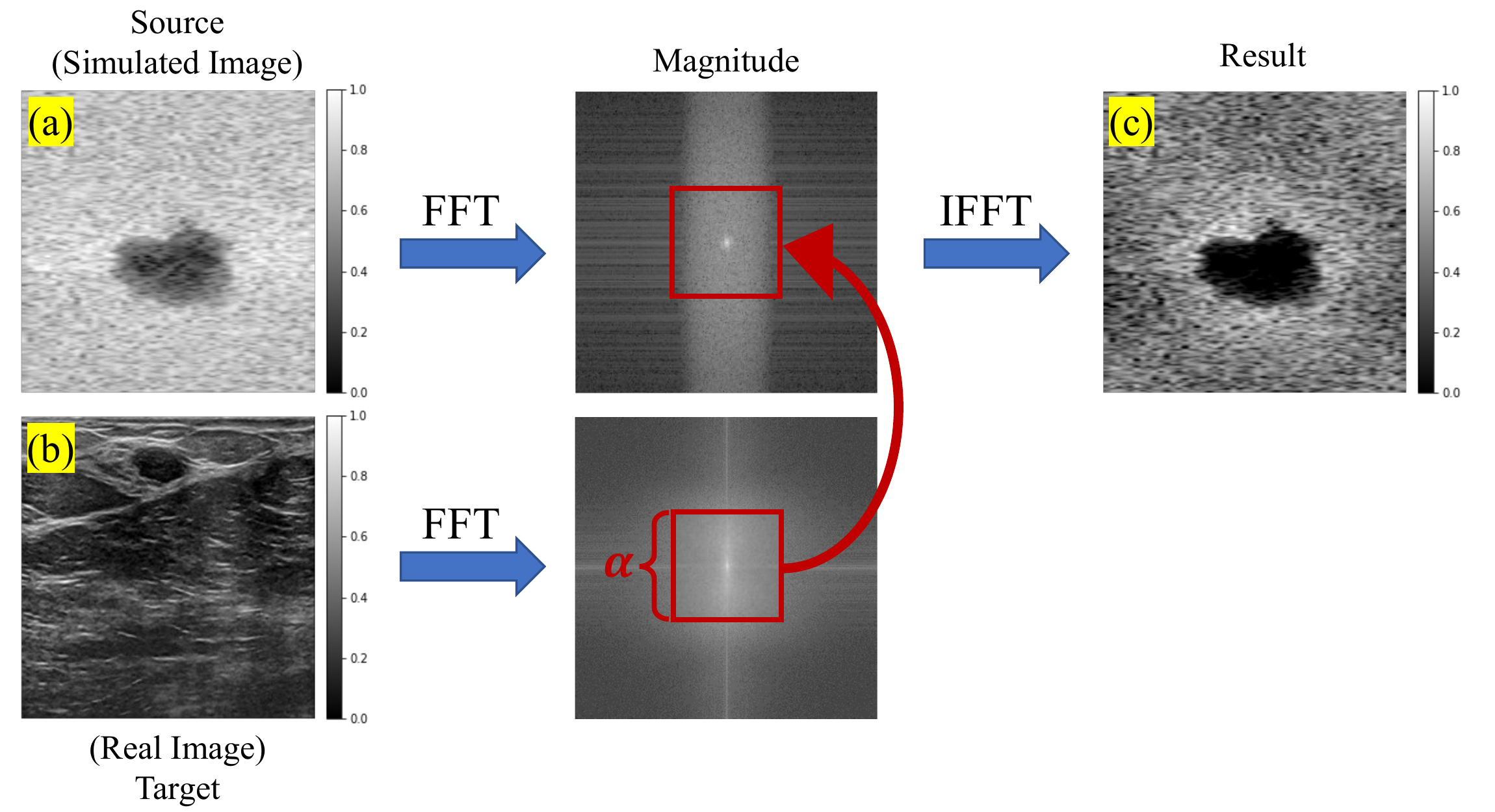}
	\caption{The FDA method takes the FFT of simulated and real images, which belong to source and target distributions, respectively. Then it replaces the magnitude of the low-frequency spectrum of the simulated image with the real one. Finally, it obtains the output by taking the IFFT from the modified simulated image. (a) A synthetic ultrasound image, which is simulated using Field II and belongs to the source distribution. (b) A real ultrasound image, which belongs to the target distribution. (c) The output, which seems closer to the target distribution.}
	\label{fig1}
\end{figure*}

\noindent Finally, given a pair of simulated and \textit{in vivo} images, the FDA method can be formalized as:
\begin{flalign}
I_{s \rightarrow t} = \mathcal{F}^{-1}(M_\alpha\boldsymbol{\cdot}\mathcal{F}_M(I_t)+(1-M_\alpha)\boldsymbol{\cdot}\mathcal{F}_M(I_s), \mathcal{F}_P(I_s))
\label{eq3}
\end{flalign}
\noindent where $\alpha \in (0,1)$. In this work, we set $\alpha=0.014$.
Fig. \ref{fig1} illustrates the FDA method. It shows (a) a simulated image $I_s$, and (b) a real ultrasound image $I_t$ from the \textit{in vivo} dataset. After taking the FFT of both images, the magnitude of the low-frequency spectrum of the simulated image has been replaced with the real one. Finally, by taking the IFFT, the output $I_{s \rightarrow t}$ has been obtained.

\subsection{Network Architecture and Training Strategy}
We used a vanilla U-Net \cite{Ronneberger2015} to evaluate the performance of the FDA method on ultrasound images for a segmentation task. The network was proposed specifically for biomedical image segmentation, where the number of available annotated samples for training is limited. It is composed of an encoder followed by a decoder, where skip connections are also adapted to concatenate low-level features of the encoder with high-level ones in the decoder.

The sigmoid function was chosen as the activation function of the output layer, and the learning rate and batch size were $1\times 10^{-4}$ and 16, respectively. The AdamW \cite{Loshchilov2019}, a variant of Adam \cite{Kingma2015}, with a weight decay of $10^{-2}$ was exploited as the optimizer.

We used Dice similarity coefficient (DSC) for evaluating the segmentation performance, and the loss function was also defined based on this metric, which quantifies the area overlap between the ground truth and predicted masks:
\begin{flalign}
DSC(S,\hat{S}) = \frac{2 \left |S \cap \hat{S} \right |+ \varepsilon}{\left |S \right |+ \left | \hat{S} \right | + \varepsilon}
\end{flalign}
where $\varepsilon$ is a small number that prevents numerical instability for small masks. For each epoch of training or fine-tuning, the model weights were stored only when the validation loss had been improved, and finally, the best weights were used for testing. Experiments were implemented using the PyTorch package \cite{Paszke2019}, and training was performed on an NVIDIA TITAN Xp GPU with 12 GB of memory.

\begin{figure}
	\centering
	\includegraphics[width=0.9999\linewidth]{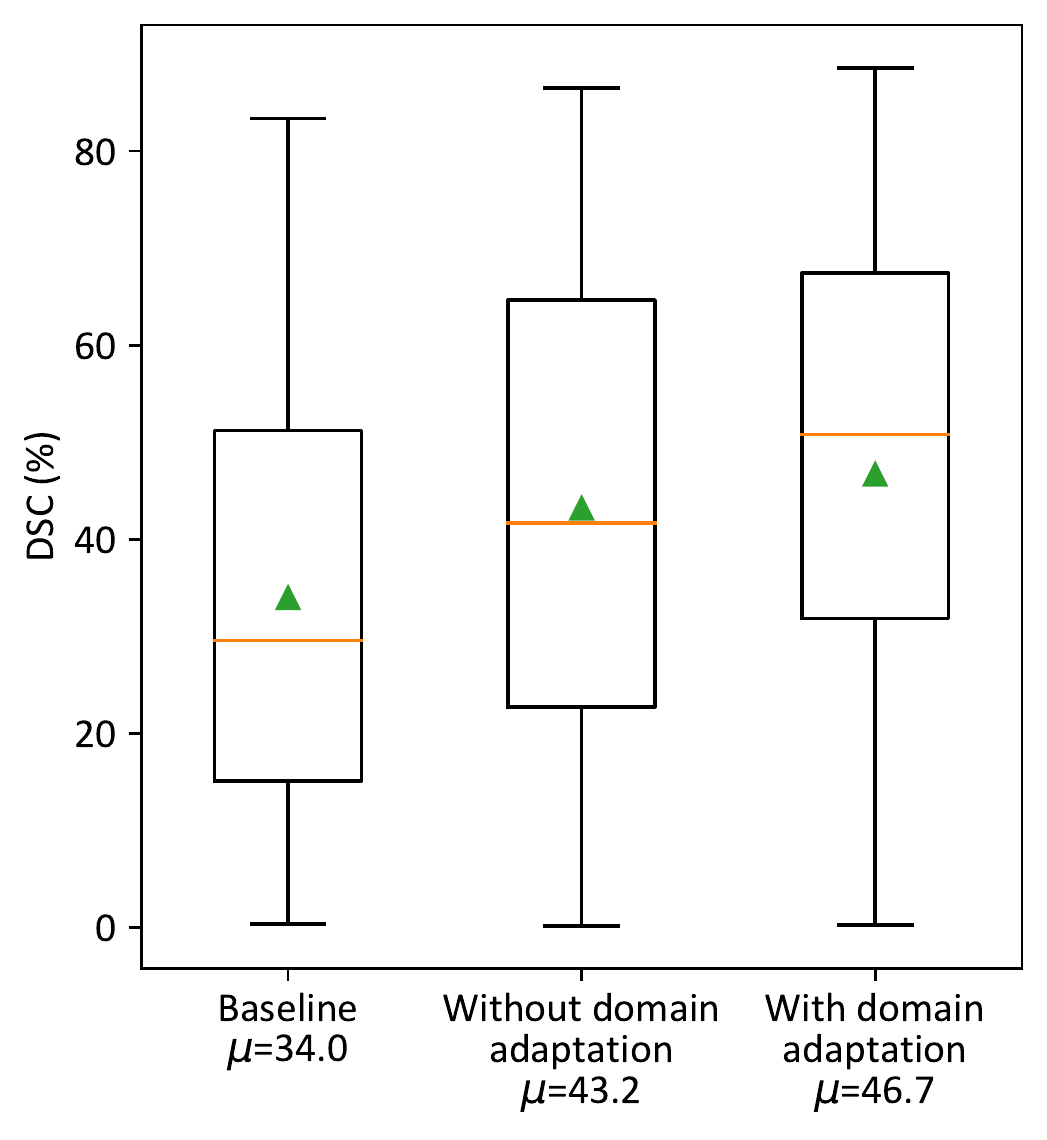}
	\caption{Quantitative comparison of DSC over the \textit{in vivo} test set.  (Left) Training the network from scratch merely using \textit{in vivo} images. (Middle) Use pre-trained weights obtained from training on simulated images without applying the FDA method. (Right) Use pre-trained weights obtained from training on simulated images with applying the FDA method. The triangle and horizontal line represent the mean and median, respectively.}
	\label{fig2}
\end{figure}

\section{Results}
To assess the effect of the FDA method for mitigating the domain shift problem, we conducted three different experiments. In the first experiment, labeled as the baseline, a network was trained merely using 20 training samples of the \textit{in vivo} dataset for 200 epochs.
In the next experiment, first, the network was trained using 800 training samples of the simulated dataset without applying the FDA method for 150 epochs. Then it was fined-tined using 20 training samples of the \textit{in vivo} dataset for 50 epochs.
Finally, as the third experiment, we applied FDA to the simulated images and repeated the previous experiment. To apply the FDA method, for each training simulated image and at each iteration, the target image was randomly chosen from 40 images in the training and validation sets of the \textit{in vivo} dataset. However, we used a fixed set of target images for applying the FDA method on the validation samples. Since the main purpose of using the validation set was to find and save the best model across different epochs, injecting randomness caused by choosing random target images was not desired. 

Fig. \ref{fig2} illustrates the DSC results for 123 test set images of the \textit{in vivo} dataset for the aforementioned three experiments. As we expected, the baseline method led to the lowest DSC due to training on only 20 real ultrasound images without taking advantage of pre-training on simulated images. The second experiment achieved a better performance by pre-training on simulated data and using \textit{in vivo} images for fine-tuning. However, it suffered from the domain shift problem, where there was a high discrepancy between the distribution of the simulated data and the \textit{in vivo} data. The third experiment showed a 3.5\% improvement in mean DSC, obtained by applying the FDA method.

\section{Conclusion}
In conclusion, we claimed that important differences between simulated and real ultrasound data are low-frequency in nature. We employed the FDA method, which replaces the magnitude of the low-frequency spectrum of a synthetic image with a real one to tackle the issue of domain shift. For the first time, we exploited the FDA method in segmentation of ultrasound images, and more generally in ultrasound imaging. We demonstrated that applying this fast and simple method on simulated ultrasound data can improve the mean DSC as high as 3.5\% compared to using simulated data without applying any domain adaptation method.

\section*{Acknowledgment}
The authors would like to thank Natural Sciences and Engineering Research Council of Canada (NSERC) for funding. We thank NVIDIA for donating the GPU.

\bibliographystyle{IEEEtran}
\bibliography{bibliography}

\end{document}